\documentclass{ws-procs9x6}
\def\be{\begin{eqnarray}&&} \def\ee{\end{eqnarray}}
\def \nonu {\nonumber \\&&}

\def\Psla{\slash \! \! \! \!}

\begin{document}

\title{Electromagnetic form factors of the nucleon 
in spacelike and timelike regions}

\author{J. P. B. C. de Melo}
\address{Centro de Ci\^encias Exatas e Tecnol\'ogicas,
  Universidade Cruzeiro do Sul, 08060-070,  S\~ao Paulo,Brazil}

\author{T. Frederico}
\address{ Dep. de F\'\i sica, ITA,
CTA, S\~ao Jos\'e dos Campos,
S\~ao Paulo, Brazil}

\author{E. Pace}

\address{Dipartimento di Fisica, Universit\`{a} di Roma "Tor Vergata" 
and
Istituto Nazionale di Fisica Nucleare, Sezione Tor Vergata, Via della Ricerca
Scientifica 1, I-00133 Roma, Italy}

\author{S. Pisano}

\address{Dipartimento di Fisica, Universit\`{a} di Roma "La Sapienza",
P.le A. Moro 2, I-00185 Roma, Italy}

\author{G. Salm\`e}

\address{Istituto Nazionale di Fisica Nucleare, 
Sezione Roma I, P.le A. Moro
2, I-00185 Roma, Italy}

\begin{abstract}
An approach for a unified description of the nucleon electromagnetic 
form factors in spacelike and timelike
 regions is presented.
The main ingredients of our model are: $i)$ a Mandelstam formula for the 
matrix elements of the nucleon electromagnetic current; $ii)$ a
$3$-dimensional reduction of the problem on the Light-Front performed
within the so-called {\tt Propagator Pole Approximation} ({\bf PPA}),
which consists in disregarding
the analytical structure of the Bethe-Salpeter 
amplitudes and of the quark-photon vertex function in the
integration over the minus components of the quark momenta;
$iii)$ a dressed photon vertex in the $q\bar{q}$
channel, where the photon is described
by its spin-$1$, hadronic component.
\end{abstract}
%
%
% ------------------- INTRODUCTION --------------------------------------
%
%

%\keywords{Style file; \LaTeX; Proceedings; World Scientific Publishing.}

\bodymatter

\section{Introduction} 
The electromagnetic form factors of the nucleon represent a powerful tool
to investigate the nucleon structure.
In particular, a detailed knowledge of the responses of the nucleon to an
electromagnetic probe can give useful information on the
quark distributions inside the hadron state, together with an insight
on the role played by the non-valence Fock components.
Presently, the experimental investigation of the nucleon electromagnetic form
factors poses various problems in both the spacelike (SL) and timelike (TL) regions.
In particular, in the SL region the most striking
observation is the discrepancy in the results obtained for the
proton ratio $G_E^p(Q^2)\mu_p/G_M^p(Q^2)$ between the data collected by
the Rosenbluth separation method and by
the polarization transfer one \cite{JLAB}.
As to the TL region, 
a sizable difference occurs between the
 theoretical expectations from perturbative QCD \cite{Karliner} - that
 predicts, at the threshold, a ratio 
$
\left (\frac{G^n_{M}}{G^p_{M}}\right)^2 \approx \left(\frac{q_d}{q_u}\right)^2
\approx 0.25 
$
- and the experimental data by Fenice \cite{Fenice}, where 
$G^n_{M} \approx G^p_{M}$.
From such a perspective, a unified investigation of both SL and
TL regions is well motivated.\\
In order to carry on the analysis for both these regions in the
framework of the same Light-Front model,
a reference frame with a non-vanishing plus component $q^+\neq 0$ is needed,
otherwise no pair-production mechanism - fundamental in the 
TL region - is allowed.
In view of this, we will adopt a reference frame where ${\bf q}_{\perp}=0$ and
$q^+ = \sqrt{|q^2|}$.
%
%
%
%
%
% ------------------- THE MODEL --------------------------------------
%
%
\section{The Model} 
The starting point of our model is the
covariant formula {\it \'a la} Mandelstam \cite{mandel}
for the nucleon electromagnetic current.
In the SL region, where the process under investigation is the scattering $e^-
N \rightarrow e^-N$, we describe the matrix elements of the {\it macroscopic}
current
%
%  ---------------------- Scattering matrix element: MACROSCOPIC ---------------------------
%
\be
\label{scatt_matr_MAC}
\langle N; \sigma', P'_{N}|j^\mu~|P_N,\sigma;N\rangle=\nonu
=\bar U_N(P'_{N},\sigma')\left [-F_2(Q^2) { {P'_{N}}^\mu +{P_N}^\mu \over 2m_N}
+\left (F_1(Q^2)+F_2(Q^2)\right )\gamma^\mu\right] U_N(P_N,\sigma)
\nonu
\ee
where the Dirac ($F_1$) and Pauli ($F_2$) nucleon form factors are present,
%
%  ---------------------- Scattering matrix element: MICROSCOPIC ---------------------------
%
by the following {\it microscopical} approximation:
\be
\langle  \sigma',P'_{N}|j^\mu~|P_N,\sigma \rangle =
3~N_c ~ \int {d^4k_1 \over (2\pi)^4}\int {d^4k_2 \over (2\pi)^4} \times \nonu
 Tr_{\tau_{(1,2)}} ~Tr_{\tau_{(3,N)}}~Tr_{\Gamma_{(1,2)}}~
\left \{~\bar \Phi^{\sigma'}_N(k_1,k_2,k'_3,P'_{N})~S^{-1}(k_1)~S^{-1}(k_2)~{\cal
    I}^\mu _3~ \times
\right. \nonu \left.
 ~\Phi^\sigma_N(k_1,k_2,k_3,P_N)\right \} ~.
\label{scatt_matr_MIC}
\ee
Analogously, in the TL region we describe the annihilation matrix
element $\langle N~\bar N|j^\mu|0\rangle$: 
%
%  ---------------------- Annihilation matrix element: MACROSCOPIC ---------------------------
%
\be
\label{annih_matr_MAC}
\langle N; \sigma',P_N|j^\mu~|-P_{\bar{N}},\sigma;N\rangle=\nonu
= \bar U_N(P_N,\sigma')~\left [-F_2(q^2) { {P_N}^\mu -{P_{\bar{N}}}^\mu \over 2m_N}
+\left (F_1(q^2)+F_2(q^2)\right )~\gamma^\mu\right]~ V_N(P_{\bar{N}},\sigma)
\nonu
\ee
%
%  ---------------------- Annihilation matrix element: MICROSCOPIC ---------------------------
%
by a {\it microscopical} approximation given by:
\be
\label{annih_matr}
\langle N; \sigma',P_N|j^\mu~|-P_{\bar{N}},\sigma;N\rangle=
3~N_c  \int {d^4k_1 \over (2\pi)^4}\int {d^4k_2 \over (2\pi)^4} \times \nonu
 Tr_{\tau_{(1,2)}} ~Tr_{\tau_{(3,N)}}~Tr_{\Gamma_{(1,2)}}~
\left \{\bar \Phi^{\sigma'}_N(k_1,k_2,k'_3,P_N)~S^{-1}(k_1)~S^{-1}(k_2)~{\cal
    I}_3^\mu~ \times
\right. \nonu \left.
\Phi^\sigma_{\bar{N}}(k_1,k_2,k_3,-P_{\bar{N}})\right \} ~.
\ee
In the previous expressions, 
$S(k)={(\Psla k +m)/ (k^2 -m^2 +\imath~ \epsilon)}$
is the quark  propagator, the quantity $\Phi^\sigma_N(k_1,k_2,k_3,P_N)$, with
its Dirac-conjugate $\bar \Phi^{\sigma'}_N(k_1,k_2,k'_3,P'_N)$, represents the nucleon
Bethe-Salpeter amplitude ({\it BSA}), and ${\cal I}^\mu_3(k_3, q)$
indicates the $q$-$\gamma$ vertex function, given by:
%
% -------------------------- Photon vertex function ------------------------------
%
\be
{\cal I}^\mu_3(k_3, q)=~\left ({1 +\tau_z \over 2} \right )~{\cal I}^\mu_u(k_3, q)~ 
+~\left ({1 -\tau_z \over 2} \right )~{\cal I}^\mu_d~(k_3, q)=\nonu
={{\cal I}^\mu_u~+~{\cal I}^\mu_d \over 2}~+~\tau_z~{{\cal
    I}^\mu_u~-
~{\cal I}^\mu_d 
  \over 2}={\cal I}^\mu_{IS}(k_3, q) +\tau_z
{\cal I}^\mu_{IV}(k_3, q)
\label{phot-vert}
\ee
%
% --------------------------------- tracce ----------------------
%
where ${\cal I}^\mu_u$ (${\cal I}^\mu_d$) indicates the $u$ ($d$) quark
contribution,  while
${\cal I}^\mu_{IS}(k_3, q)$ (${\cal I}^\mu_{IV}(k_3, q)$)
is the isoscalar (isovector) contribution.\\
%
%
% ------------------------ VERTEX FUNCTIONS AND BETHE-SALPETER AMPLITUDES ----------------------------
%
%
%
% --------------------------------- lagrangian ------------------------
%
To introduce a proper Dirac structure 
for the nucleon {\it BSA}, we describe the $qqq$-$N$ 
interaction
through an  effective Lagrangian, which  represents an isospin-zero and 
 spin-zero coupling for the (1,2) quark pair, as in Ref. [\refcite{nua}] but with $\alpha = 1$. 
Then, the nucleon {\it BSA} is approximated as follows
%
% ------------------------- nucleon BS -------------------------
%
\be
\Phi^{\sigma}_N(k_1,k_2,k_3,p) 
=\left [~S(k_1)~\imath \tau_y ~ \gamma^5 ~ S_C(k_2)C ~ S(k_3)~+
~S(k_3)~\imath \tau_y ~ \gamma^5 ~S_C(k_1) \right. \nonu \left.
\times C ~ S(k_2)~+ ~S(k_3)~\imath \tau_y ~ \gamma^5 ~S_C(k_2)C~S(k_1)~
 \right ] 
 ~\Lambda(k_1,k_2,k_3)~\chi_{\tau_N}~U_N(p,\sigma)
 \nonu
 \label{ampli1}
\ee
where: i) $\Lambda(k_1,k_2,k_3)$ describes  the symmetric momentum dependence of the
vertex function on the quark momentum variables, $k_i$,  
ii) $U_N(p,\sigma)$ is the nucleon  Dirac spinor,
iii) $\chi_{\tau_N}$ the nucleon isospin state and iv) $S_C(k)$
is the  charge conjugate quark propagator.\\
%
%
% --------------------- 3-dimensional reduction -----------------------------
%
%
%
% ------------------------------ PROCESS FIGURES (SL and TL) ------------------------------
%
\begin{figure}[htb]
 \begin{center}
\includegraphics[width=10.cm]{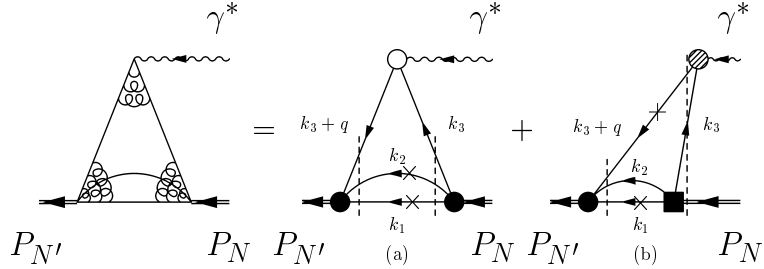}
\end{center}
\caption{Diagram (a) : valence, triangle contribution  ($0 < k_{i}^+ < P^+_{N}$,
$0 < k_{3}^+ + q^+ < P^+_{N^\prime}$).
 Diagram (b) : non-valence contribution  ($0 > k_3^+ > - q^+$). The symbol $\times$ 
 on a quark line indicates a quark on the mass shell, {\it i.e.}
$ ~ k^-_{on} = (m^2 + k^2_{\perp})/k^+$. (After Ref. [\refcite{fb18}]).
\label{SL_proc}}
\end{figure}
\begin{figure}[htb]
\begin{center}
\includegraphics[width=10.cm]{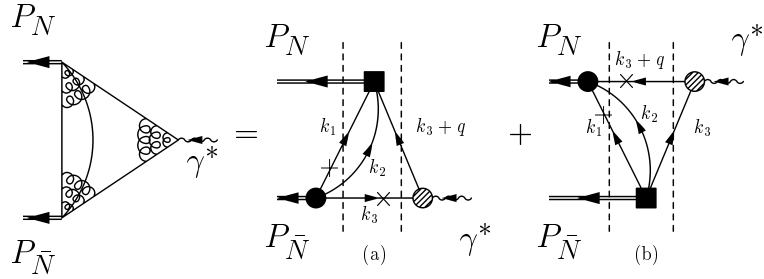}
\end{center}
\caption{Diagrams contributing to the nucleon EM form factors in the 
timelike region. The solid circles represent the on-shell amplitudes, 
the solid squares represent the non-valence ones and the 
shaded circles represent the dressed photon vertex.
\label{TL_proc}}
\end{figure}
In order to apply phenomenological approximations inspired by the
Hamiltonian language to our model, we need to project the matrix elements on
the Light-Front.
To this end, we integrate over the minus components of the quark momenta. 
We assume a suitable  fall-off for the functions 
$\Lambda(k_1,k_2,k_3)$ and $\Lambda(k_1,k_2,k'_3)$ appearing in the nucleon
{\it BSAs}, 
to make finite the four dimensional integrations.
Furthermore, we assume that the singularities of 
   $\Lambda(k_1,k_2,k_3)$ and $\Lambda(k_1,k_2,k'_3)$ 
give a negligible contribution to the integrations on {$k_1^-$} and on {$k_2^-$} and
 then these integrations are performed taking into account only the  poles 
of the quark propagators.\\
%
% -------------- kinematical regions -------------------------------------
%
The $k^-_i$ integrations automatically single out two kinematical
regions, namely a {\it valence} region, given by a triangle
process (Fig. 1 (a)),
with the spectator quarks  on their
 mass shell and both the initial and the final nucleon vertexes in the valence sector,
and the {\it non-valence} region (Fig. 1 (b)), where the $q \bar q$ pair production appears and only the final 
nucleon vertex is in the valence sector. This latter term can be seen as a 
higher Fock state
contribution of the nucleon final state to the form factors.
%
% ----------------------- quark-photon vertex ---------------------------------
%
According to this kinematical separation, 
both the isoscalar and the isovector part of the quark-photon vertex ${\cal I}^\mu_3$
contain a purely bare valence contribution and a contribution corresponding to the pair
production ($Z$-diagram),
which can be decomposed in a
bare, point-like term and a vector meson dominance (VMD) 
term (Fig. \ref{fig:qdress}):
\be
{\cal I}^\mu(k,q,i) = {{\cal N}_{i}}~\theta(p^+-k^+)~\theta(k^+)
\gamma^\mu+ \nonu
+ \theta({q}^+ + k^+)~
\theta(-k^+) \left [{Z_b ~{\cal N}_{i}} ~\gamma^\mu + Z_{V} \Gamma^\mu(k,q,i)\right]
\label{vert} 
\ee
 with $i = IS, ~IV$ and ${\cal N}_{IS}=1/6$, ${\cal N}_{IV}=1/2$. 
The first term in Eq. (\ref{vert}) is the bare coupling of the triangle contribution, 
while $Z_b , ~Z_{V} $ are renormalization constants to be determined from
 the phenomenological analysis of the data. As a consequence the SL form 
factors $F^N$ are the sum of a valence term, $F^N_{\Delta}$,
plus a non-valence one, $F^N_{\cal{Z}}$.
%
%
%
%
%
% ------------------------------ DRESSED PHOTON FIGURE ------------------------------
%
\begin{figure}[hbt]
  \begin{center}
\includegraphics[width=11.5cm]{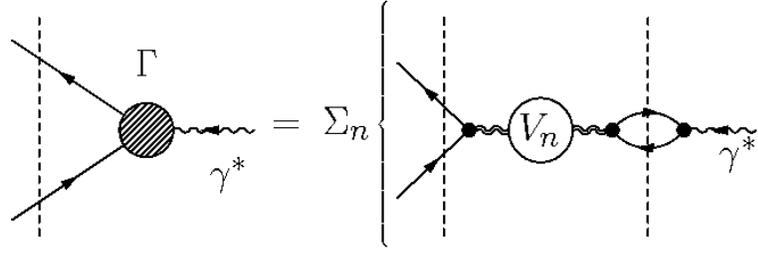}
\caption{Diagrammatic analysis of the VMD contribution to the quark-photon vertex, 
$\Gamma^\mu(k,q)$. (After Ref. [\refcite{plbpion}]).
\label{fig:qdress}}
\end{center}
\end{figure}
%
%
% ---------------------- PHENOMENOLOGICAL APPROXIMATIONS ---------------------------
%
%
\section{Phenomenological approximations}
In order to overcome the lack of solutions for the $4$-dimensional
Bethe-Salpeter equation in the baryon case, we insert in our model some
phenomenological approximations.
In particular, we test an Ansatz for the momentum dependent part
$\Lambda(k_1,k_2,k_3)$
of the nucleon BSA in Eq. (\ref{ampli1}), while the VMD description
of the photon hadronic component, that appears in the quark-photon vertex,
will be described by the eigenstates of a relativistic, squared mass operator
introduced in Ref. [\refcite{FPZ02}].\\
In more details, the various amplitudes appearing in Figs. \ref{SL_proc},
\ref{TL_proc}
will be described as follows:
\begin{itemize}
\item for the solid circles, representing the valence, on-shell amplitudes, we adopt a
  power-law Ansatz {\it \'a la} Brodsky- Lepage \cite{brds_lep};
\item for the solid squares, representing the non-valence, off-shell amplitudes, we adopt
  a phenomenological form where the correlation between the spectator quarks is implemented;
\item for the empty circle, representing a bare photon, a pointlike vertex, 
$\gamma^{\mu}$, is
  used;
\item for the shaded circles, representing a dressed photon, we use the sum of a bare term
  and a microscopical VMD model \cite{plbpion}.
\end{itemize}
The actual forms are the following.
For the on-shell nucleon amplitude, we have:
%
% ----------- on-shell nucleon amplitude ----------------------
%
\be
\Psi_N(k^+_1,{\bf k}_{1\perp},k^+_2,{\bf k}_{2\perp},P_N) =
P_N^+ \frac{\Lambda(k_1, k_2, k_3)}{m_N^2- M_0^2}
\sim {{\cal N}~ P_N^+ ~ {\cal F}(\xi_1, \xi_2, \xi_3) \over \left[\beta^2 +
    M^2_0(1,2,3)\right]^3} ~~~~
\label{mom_dep_ON}
\ee
where ${\cal F}(\xi_1, \xi_2, \xi_3)$ is a symmetric scalar function depending on the
momentum fraction carried by each constituent and $M_0(1,2,3)$ is the light-front free mass 
for the three-quark system. At the present stage we take  ${\cal F}(\xi_1, \xi_2, \xi_3)=1$.
As to the non-valence amplitude, it is described by:
%
% ----------- on-shell nucleon amplitude ----------------------
%
\be
\label{mom_dep_OFF}
\Lambda(k_1,k_2,k_3)\sim \nonu
\sim {{\cal G}(\xi_1, \xi_2, \xi_3) \over \left [\beta_{off}^2+M^2_0(1,2)\right]^p}~\left \{ {1 \over \left
[\beta_{off}^2+M^2_0(3',2)\right]} +{1 \over \left
[\beta_{off}^2+M^2_0(3',1)\right]}\right \}~~~~
%\nonu
\ee
where $M_0(i,j)$ is the light-front free mass for a two-quark system and
${\cal G}(\xi_1, \xi_2, \xi_3)$ is  a symmetric function of $\xi_i$.
In this preliminary work ${\cal G}(\xi_1, \xi_2, \xi_3)=1$ and $p=2$ are chosen.
As to the $q$-$\gamma$ vertex, the term $\Gamma^\mu(k,q,i)$ in Eq. (\ref{vert})
is obtained through the same microscopical VMD model
already used in the pion case with the same VM eigenstates \cite{plbpion}.
Eventually, given the simplicity of the forms adopted for the
nucleon amplitudes, a further dependence on the momentum transfer is 
introduced in the SL region through a modulation function $
f(Q^2) = (1+aQ^2)/(1+bQ^2)
$,
that parameterizes all the effects not included in 
the na\"ive amplitudes we are using. 
This function acts in the following way on the total
form factors $F^N$: $F^N(Q^2)= F^N_{\Delta} f^2(Q^2) + F^N_{\cal{Z}}f(Q^2)$.
%
%
%
%
%
%
%
%
% ------------------ RESULTS ----------------------------------------
%
%
\section{Results and perspectives}
%
%
%

%
%
%
% --------------------------- FIGURE CON I RISULTATI SPACELIKE --------------------------
%
%
\begin{figure}
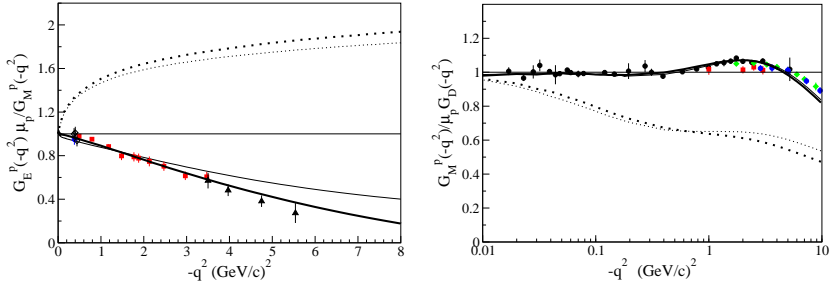


\vspace {.1 cm}
%
% -------------------------- proton figure ----------------------------------
%
\parbox{5.3cm}{\includegraphics[width=5.3cm]{ThGepmupGmp.eps}}
$~~$\parbox{5.3cm}{\includegraphics[width=5.3cm]{ThRGMpGD2.eps}}
\vspace{0.05 cm}
\caption{{\bf Left panel}. The ratio  $G^p_E(Q^2) \mu_p/G^p_M(Q^2)$ vs $Q^2$.
  Thin (thick) solid line: full calculation for a quark mass $m_q= 190$ ($m_q= 200$) MeV, 
  corresponding 
  to the sum  of all the contributions to 
  $G^p_E(Q^2) $ and $G^p_M(Q^2)$, {\it i.e.} triangle plus pair production terms. 
  Thin (thick) dotted line: triangle elastic contribution to $G^p_E(Q^2) $ and $G^p_M(Q^2)$
  for a quark mass $m_q= 190$ ($m_q= 200$) MeV.  
  {\bf Right panel}. $G^p_M(Q^2)/\mu_p G_D(Q^2)$ vs $Q^2$, with  the same notations as 
  in the left panel, and  $ G_D(Q^2)=1/(1+Q^2/0.71)^2$. For experimental data
  see Ref. [\refcite{JLAB2}].
  \label{fig_SL_p_res}}
\end{figure}
%
% -------------------------- neutron figure ----------------------------------
%
\begin{figure}
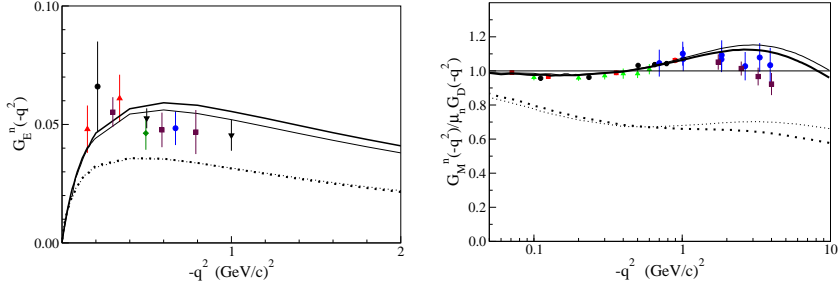


\vspace {.1 cm}
\parbox{5.3cm}{\includegraphics[width=5.3cm]{ThGEn.eps}}
$~~~$\parbox{5.3cm}{\includegraphics[width=5.3cm]{ThRGMnGDext.eps}}

\vspace{0.05 cm}
\caption{{\bf Left panel}.   $G^n_E(Q^2) $ vs $Q^2$.
  Solid line: full calculation, 
  corresponding to the sum of all the contributions to 
  $G^n_E(Q^2) $, {\it i.e.} triangle plus pair production terms. 
  Dotted line: triangle elastic contribution to $G^n_E(Q^2)$.  
  {\bf Right panel}. $G^n_M(Q^2)/\mu_n G_D(Q^2)$ vs $Q^2$
  and $ G_D(Q^2)=1/(1+Q^2/0.71)^2$. The notation is the same as in Fig. \ref{fig_SL_p_res}.
  For experimental data see
  Ref. [\refcite{JLAB2}].
  \label{fig_SL_n_res}}      
\end{figure}
%
%
%
%
% --------------------------- FIGURE CON I RISULTATI TIMELIKE ---------------------------
%
%
%
% -------------------------- proton figure TL ----------------------------------
%
\begin{figure}
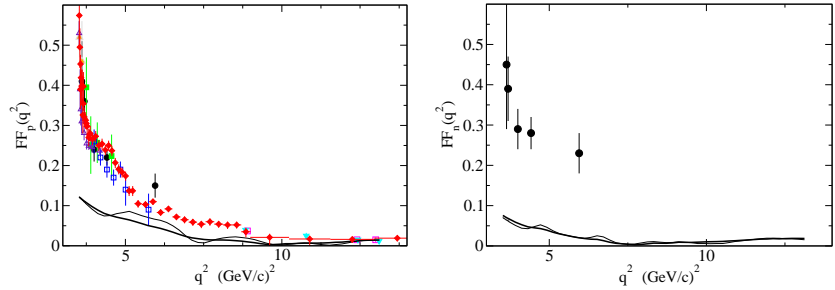

\begin{center}

\vspace{0.05cm}
\parbox{5.3cm}{\includegraphics[width=5.3cm]{ThFFpTL_271106.eps}}
$~~$\parbox{5.3cm}{\includegraphics[width=5.3cm]{ThFFnTL_271106.eps}}
\vspace{0.1 cm}
\caption{{\bf Left Panel}. Proton effective form factor  in the timelike region.
  Thin (thick) solid line: calculation for a quark mass $m_q= 190$ ($m_q= 200$) MeV.
  For the experimental data see Ref. [\refcite{babar}] and Refs. therein.
  {\bf Right Panel}. Neutron effective form factor in the TL
  region, with the same notation of the Left Panel. 
  Data by Fenice Collaboration\cite{Fenice}.
  \label{fig:FFTL}}
\end{center}
\end{figure}

The preliminary results obtained with our model are reported in Figs. 
\ref{fig_SL_p_res}, \ref{fig_SL_n_res} and \ref{fig:FFTL}.
They show clearly the relevant role played by the pair
production process - and then by the non-valence components -
for the nucleon form factors.
In particular, the interplay between the valence and non-valence contributions generates
the possibility to have a zero in the proton electric form factor.
A reasonable description of the experimental data in the SL region is obtained,
with a modulation factor $f(Q^2)$ which grows from 1 to 1.08, at most.
As to the TL region, our preliminary results reproduce
 the behavior of the experimental data,  but for a scale factor.\\
%
%
%
%\section{Perspectives}
%
%
%
Various improvements can be applied to our model.
First of all, a more elaborate form for the nucleon non-valence amplitude - that
is tested for the first time in this model - could allow for a better
description of the experimental data in the TL region. 
Another improvement can derive from a better description of the
quark-photon vertex, and in particular for the VMD approximation.
A new, fully covariant VMD model is presently under investigation.%

\end{document}